# Acousto-optic holography for micrometer-scale grid patterning of amplified laser pulses with single-pulse accuracy

Walther Akemann and Laurent Bourdieu

Institut de Biologie de l'ENS (IBENS), École Normale Supérieure, CNRS, INSERM,

Université PSL, 75005 Paris, France

**Abstract**

Many optical systems use acousto-optics for fast steering of light. However, acousto-optic (AO) light diffraction permits manipulation and shaping of light fields in a much broader sense including holography. While acousto-optic devices operate at incontestable speed, they are one dimensional (1D) diffraction devices depending on analog serial data transfer and as such have a mode of use distinguished from common 2D spatial light modulators. Combination with an amplified laser source allows to clock the serial buildup of AO diffraction patterns for avoidance of phase aberration in the output beam due to acoustic wave progression. Pseudo 2D holography is established with two crossed AO deflectors for biaxial spatial modulation of the input phase, and optionally also of its amplitude. In this configuration, acousto-optics permits light patterning into micrometer-scale 2D point grids with the possibility to switch the pattern between individual laser pulses. We describe algorithms for deriving AO holograms for either pure phase modulation or for a co-modulation of amplitude. Computationally reconstructed phase holograms show focal patterns in presence of side lobes. Side lobes are entirely suppressed by amplitude modulation, though at the expense of reduced optical power transmission.

**Introduction**

Optical holography uses the spatial amplitude and phase distribution of electromagnetic light fields as a carrier to encode, store and retrieve information[1]. Making use of fast-switching pixelated modulators able to reproduce digitally-stored two-dimensional (2D) holograms in rapid succession, holography is serving modern optical technology in many areas, including for data display, data storage, 3D imaging, lithography, additive manufacturing, cryptography, remote sensing and the visual arts[2–12]. Generally, digital holography combines fast parallel processing of high data loads with high input-to-output transfer efficiency, as data may be encoded partially, or even entirely, as modulated phase instead of intensity. In addition, using classical electrodynamic theory, holograms and/or their reconstruction may be obtained by digital computation for full compatibility and seamless interfacing of optical and digital data processing[13].

Practically, holography applications often have to compromise between update speed and optical resolution. Spatial light modulators (SLMs) using nematic liquid-crystal arrays for the display of phase holograms can provide high pixel resolution (1-5 MP; mega pixel) at moderate cycle time (20-100 Hz), whereas micro-electromechanical system (MEMS)-based movable mirrors have fast cycle times (10-200 kHz) at modest pixel resolution (0.1-4 kP; kilo pixel). Less employed are acousto-optic (AO) devices in spite of fast cycle time (20-100 kHz) for phase and amplitude modulation, as the modulation is only one dimensional (1D) at intermediate line resolution (0.05-0.4 kP). Acousto-optic light modulation, on the other hand, excels if high speed is most important and 1D modulation sufficient. One example are low-noise holographic displays where 2D image frames are reconstructed from sets of 1D line holograms through an in-phase linear scan in orthogonal direction, still fast enough to permit temporal filtering of speckle noise at video rate[14]. In functional scanning microscopy, as another example, fastest data acquisition is achieved in random-access mode using fast AO light modulation[15,16], including for 3D random-access[17,18] in combination with holographic light patterning[19,20] and fast adaptive correction of light scattering aberrations[21].

In the following we outline the operational principles and mode of use of acousto-optic deflector (AOD) devices as general spatial light modulators. Using computational reconstruction, we demonstrate the difference in light patterning achieved by holography with only the phase being modulated versus combined phase and amplitude modulation. Although motivated as a contribution to holographic methods in functional bioimaging, the results may serve focal grid patterning of amplified laser pulses in other applications, in particular where fast pulse-to-pulse pattern update and random-access capability are of advantage.

**Acousto-optic 3D laser beam steering**

Acousto-optic deflectors (AODs) permit faster laser beam scanning than other existing technologies[22]. In such devices a collimated laser beam is Bragg-diffracted from a travelling longitudinal density wave[23] in an optical crystal, with the density wave carried by acoustic long-wavelength phonons close to the center of the crystal first Brillouin zone. The devices operate in a strictly serial mode. An externally supplied radio signal is transduced into a propagating sound wave through a piezoelectric element, resulting in a one dimensional spatial phase grating across the crystal aperture as a retarded replica of the time-dependent input signal[24]. Optimal diffraction efficiency in first order is achieved through resonant phase match between the wavevector of the acoustic carrier wave and the wavevector difference of the incoming and diffracted beams. For large acoustic bandwidth, high performance AODs use a particular resonance in biaxial birefringent crystals, where an extraordinary input beam is diffracted into an ordinary output beam such that their wavevector difference is cutting tangentially through the ordinary index surface of the crystal[25]. This tangential resonance ensures quasi-phase matched conditions over a large acoustic bandwidth and thereby enables sideband frequency modulation for deflection tuning around the Bragg angle, the diffraction angle of the carrier wave[24] (Fig. 1a). In this way, AODs made from chalcogenide crystals, like $TeO_2$, may offer an angular tuning range of 10 to 50 mrad with 60 to 80 % efficiency in first order[22]. AODs also have a lensing property, which refers to the phenomenon that rapid linear acoustic frequency change will create a chirped diffraction pattern mimicking a diffraction cylindrical lens (Fig. 1b) with its focal length scaling reciprocally with the chirp rate[26]. The combination of tunable beam deflection and lensing provides for 3D beam scanning. In scanning microscopy, in particular, the combination allows to move the beam focus under the objective freely in lateral and axial direction without mechanically moving parts[17]. However, as the acoustic wave continues to propagate across the crystal with constant chirp, the grating constants are constantly changing giving rise to a lateral motion of the focus at ultrasound speed, when illuminated with continuous or quasi-continuous light (Fig. 1c). This motion can be prevented by passing the beam through a second acousto-optic (AO) crystal filled with an acoustic wave of same chirp but opposite direction of propagation[27], a configuration that has been implemented in a number of applications[28–30]. However, this scheme encounters severe wavefront distortions when the chirp is reset to keep acoustic frequencies within the supported range (Fig. 1d). Unfortunately, the shorter the focal length, the more frequently the reset is required, thereby deteriorating the duty cycle as function of AO focal length[17]. Furthermore, the dual AOD configuration suffers from transmission variance because the second grating may receive wavefront input featuring local input angles different from the Bragg angle. As perhaps most consequential shortcoming, the stabilization of the emerging wavefront by the second grating breaks down for general non-linear frequency chirps, thus limiting the avail of acousto-optics for purposes of general light modulation.

**Acousto-optic spatial light modulation**

Previous work[31] has established an alternative AOD operational mode suitable to suppress phase drift in the outgoing wavefront due to AO wave progression. In this synchronous mode the AOD update cycle is time-locked to the laser pulsing such that the laser beam is diffracted from the AO grating precisely only, when the aperture is completely filled with the programmed diffraction pattern at the end of an AOD write process, thereby skipping the time of pattern buildup[31,32]. The synchronous mode is illustrated in Fig. 1e for the case of five laser pulses, individually modulated to address different focal positions in the object space under an objective (Fig. 1f). The AO write process has an intrinsic cycle time of $d_{AOD}/v_{ac}$, with $d_{AOD}$, the AOD aperture diameter and $v_{ac}$, the acoustic speed of acoustic sheer waves in TeO$_2$ (650 m/s)[24], giving rise to 12 μs and 23 μs for 8 and 15 mm aperture sizes, respectively. As the AOD cycle time defines the usable laser pulse repetition rate, the synchronous mode is generally suited for laser pulses up to 100 kHz, which makes this mode particularly efficient in combination with laser amplifiers.

While AODs thus excel in speed, they are strictly one-dimensional diffraction devices. However, pseudo two-dimensionality can be achieved with two AODs in crossed configuration. Two optically conjugated AODs receiving equal linear frequency chirp produce a spherical lens (Fig. 1g), by creating the same spherical phase on both axes (Fig. 1h), while frequency offsets to either axis produce linear phase tilts in x,y-directions (Fig. 1 h). As these phase functions are orthogonal, and thus linear independent, their linear superposition enables 3D random-access, with the 3D Cartesian point coordinates coded for by the weights attributed to tilt and defocus phases. Importantly, the synchronous mode eliminates the phase drift in the outgoing wavefront associated with drift progression of AO diffraction gratings in the case of arbitrary AO frequency chirp, which unlocks the full potential of acousto-optics for pulse-resolved spatial light modulation of high-power laser beams. One way to understand the principal is by analogy with the Shack-Hartman wavefront sensor. In the Shack-Hartman sensor, the slope of a spatial phase of a light beam is reconstructed from a measure of angular tilts of its wavefront[33]. AO light modulation can be seen as the inverse: the outgoing wavefront is synthetized to any desired shape, up to the spatial resolution limit of the AODs[34,35], by imposing appropriate local angular tilts onto the input wavefront through local AO frequency offsets (Fig. 2a). Quantitatively, the slope of local phase $\varphi_y(y)$ of the emerging wavefront at position y within the AOD aperture turns out as $d\varphi_y(y)/dy =$ $2\pi/v_{ac} \cdot f_{ac}(y)$, with $f_{ac}(y)$, the local frequency of the AO diffraction pattern[31]. A desired output wavefront is therefore obtained by frequency modulation (FM) of the AO carrier grating with the modulation function given by the slope of the desired phase. The synchronous mode, in this way, enables AODs to operate as 1D spatial light modulators with single pulse accuracy for arbitrary phase and amplitude functions, with general polynomial functions as an example (Fig. 1i). In orthonormal representation, first and second order polynomials reproduce Zernike polynomials of the same order, whereas for higher orders the functions deviate from the Zernike modes because of their imposed biaxial symmetry, with the only exception of vertical astigmatism of 4th polynomial order (Fig. 1i).

**Fast acousto-optic light patterning**

In digital holography, computer-generated holograms (CGHs) are derived from a target function using the laws of diffraction optics or, for simpler calculation, of Fourier optics. When approximating the imaging objective by an aberration-free, paraxial Fourier lens, the light field in the front focal plane of the lens is obtained as the Fourier transform of the light field in the back focal plane (Fig. 2b), also called the holographic plane[36]. A CGH is then calculated simply by inverse Fourier transformation of a given frontal focal light field back to the holographic plane. Yet, the target function only constraints the amplitude of the field in the front focal plane, whereas the phase of the field is usually not known. A suitable phase can nonetheless be found iteratively from an initial random phase by propagating the light fields back and forth between front and back focal plane, by inverse and forward Fourier transformation, under the constraint of chosen input and target amplitudes until the phases (φ and φ' in Fig. 2c) become self-consistent under this transformation (Iterative Fourier Transform Algorithm[37]). Often, however, the knowledge of the desired target amplitude pattern is itself incomplete. What is usually given is the wanted amplitude distribution in the first diffraction order but not in higher orders. Clamping higher orders to zero, on the other hand, would result in a target function incompatible with the laws of diffraction optics. One way to solve this issue is by use of an adaptive target function to enable the algorithm to learn the required higher order amplitude distribution during iteration, while keeping the first order target unchanged (Fig. 2c). This can be implemented in different ways. One possibility is to initialize side lobes with arbitrary initial values and let these evolve during iteration under suitable constraints minimizing the target error, which is defined as deviation of the reconstructed from the wanted target distribution restricted to the first diffraction order. Given the biaxial pseudo two-dimensionality of the modulator, valid holograms need to be separable with regard to their x- and y-coordinates[31], thus $\varphi(x,y) = \varphi_x(x) + \varphi_y(y)$ and $A^*(x,y) = A_x^*(x) \cdot A_y^*(y)$. The search algorithms (Fig. 2c) is therefore performed independently for one-dimensional fields aligned with either the x- and y-axes of the modulator (Fig. 2c) and the retrieved 1D functions are subsequently recombined into a 2D hologram. To demonstrate the method, we calculated phase holograms for several target patterns, as examples, and computer-reconstructed the resulting amplitude distribution in the focal plane of an imaging objective. The simulations were performed for a collimated input beam of flat intensity, 10 mm diameter and 900 nm wavelength, with the imaging objective modeled as a 8 mm Fourier lens. In the present context these values are arbitrary and only meant to illustrate the method. Fig. 2d shows the results for four target patters of one to seven equidistant points. Reconstruction of the focal amplitude distribution (A' in Fig. 2c) reproduced the position and amplitudes of the wanted target points together with higher order side lobes not included in the original target function (Fig. 2d, *left*), while the modulation bandwidth increased with the number of points to generate (Fig. 2d, *right*). The latter point also manifests through appearance of $2\pi$ phase jumps in the hologram of the seven-point target (Fig. 2d, *right*). Applying the same modulations to both axes created symmetric 2D phase holograms reconstructing into 3x3, 5x5 and 7x7 grids, respectively, of equidistant points in the focal plane (Fig. 2e). Consistent with the 1D representation (Fig. 2d), side lobes are discernible in the 2D reconstructed amplitude image and expectedly weaker in the intensity image (Fig. 2e). Applying different targets to the x- and y-axis reduced the symmetry of the 2D hologram and its reconstruction from four- to two-fold (Fig. 2f). As expected from Fourier optics, increasing the spatial frequency of a grating in the holographic plane increased the point spacing in the focal plane by the same factor (Fig. 2g). Finally, imposing non-equidistant point targets gives rise to generally asymmetric holograms of higher complexity and larger modulation depth (Fig. 2h). While these are illustrating examples, the range of principally accessible grid geometries remains constrained by the biaxiality of the pattern generator. In consequence, accessible geometries must exhibit translation symmetry along each axis and the number of total points is given by the product of the number of points on each axis. Freely selectable, on the other hand, are the distances between neighboring points, thereby also defining the overall envelope size of the pattern.

**Acousto-optic holography with amplitude modulation**

While most of currently used SLM devices are pure phase modulators, AODs are capable to represent phase and amplitude functions at the same time and with identical update speed. Combined frequency and amplitude modulation (FM/AM) augments the overall light shaping capability of the modulator. While phase holography (FM) inevitably produces higher order side lobes, albeit of low intensity (Fig. 2 d-g), the FM/AM mode is suitable to suppress higher diffraction orders altogether. In this case the hologram is calculated as before (Fig. 2c), but with the full target function as a static target (Fig. 3a). The output is a hologram which includes a phase ($\varphi$) and an amplitude (A*) function (Fig. 3a). While the FM mode (Fig. 2c) only uses the holographic phase for reconstruction, the FM/AM mode uses the amplitude as apodization filter for the input beam in post-iteration reconstruction, thereby exploiting the entire phase and amplitude information encoded in the hologram. For direct comparison Fig. 3b shows the results for the same targets as before (Fig. 2d). The reconstruction of the phase holograms alone (FM), when derived from static targets, exhibit irregular target peaks leading to significant target error (Fig. 3b, *middle*). This is an expected outcome as the algorithm is forced to compromise between two incompatible constraints, namely constant intensity of the target points and zero intensity of side lobes. This is exactly the situation that was remedied in the previous section by using an adaptive target function in the phase search in FM only mode. (Fig. 2c). In FM/AM mode, on the other hand, target peaks are well reproduced with close to zero error, comparable to the FM mode using adjustable targets (Fig. 2d), but in notable absence of higher diffraction orders (Fig. 3b, *middle*). Reconstruction of the 2D holograms (Fig. 3c) confirm the absence of peaks of higher order in the FM/AM mode (Fig. 3d, *right*). To understand how this is achieved, it is informative to assess the reconstructed phase distributions in the focal plane (Fig. 3d, *middle*) or, for better visibility, the spatial phase variance (Fig. 3d, *right*). In the phase image, the target positions show up as circular areas of constant phase or, equivalently, areas of low spatial phase variance, in both modes. In the FM only mode however, areas of constant phase also exist outside of the target zone at positions where side lobes appear in the amplitude image. Notably, these latter areas associated with higher order diffraction are missing in the FM/AM mode. The effect of amplitude modulation therefore consists in scrambling the phase of the side lobes by producing local phase variance to inhibit constructive beamlet interference at these positions and to redistribute intensity to the wanted target points. While amplitude modulation results in a focal intensity distribution free of higher diffraction orders, which could have advantage for certain applications, it comes with the price of reduced power transmission. To investigate this point, we varied the AM modulation gain between zero and one, with one corresponding to the amplitude found in Fig.3b, for the five-point pattern as example (Fig. 3e) and evaluated the target approximation error and power transmission for each gain (Fig. 3f). Target fidelity increased linearly with modulation gain, from 17 % at zero gain to zero at full gain, while higher order amplitudes decreased (Fig. 3e). As expected, the power transmission decreased linearly with increasing gain, reaching about 50 % at full gain (Fig. 3f), and would further halve, if the beam passed through a second crossed AOD performing the same modulation.

**Conclusions**

The above simulations show how to program an AO spatial light modulator to produce focal point grids of variable size and point density within the optical resolution limit of the imaging objective. Compared to other spatial phase modulators, AODs have the advantage of high update speed in match with the repetition rate of laser amplifiers, high dynamic range without $2\pi$ phase reset and high damage threshold with practically unlimited life time. Their cycle time is set by the time an acoustic signal needs to travel from the injection to the absorption face of the AO crystal[24]. Therefore, smaller crystals have faster cycle time but lower optical resolution because of a decreased diffraction aperture[34]. Although optical resolution is ultimately limited by chromatic dispersion occurring in the AODs due to diffraction of transform-limited laser pulses, dispersive aberration can to a large part be corrected for by appropriate compensation[34,35]. In the context of holography, the most important limitation remains the bi-axiality of the provided modulation. As a consequence of bi-axiality, 2D focal patterns are constraint to a bidirectional translational symmetry implying necessarily a rectangular border envelope, while number and distances of points are freely programmable. As the target points define a set of discrete acoustic frequencies, excluding spatial frequencies close to zero, the grid patterns are not sensitive to the phase drift induced by the progression of the AO diffraction pattern, different from patterns carrying frequency chirp associated with polynomial phases such as defocus (Fig. 1c). Therefore, AO holography does not necessarily require the synchronous mode. In asynchronous mode, however, switching between patterns would still create transient aberration for the duration of one cycle time (Fig. 1d). As a consequence, switch from one pattern to the next is not faster in the asynchronous than in the synchronous mode, although the asynchronous mode permits diffraction of light rom from fast-repetition rate or continuous lasers. In addition, only the synchronous mode provides the convenience of single-pulse patterning and compatibility with defocus and other higher order phases (Fig. 1i).

Holographic light patterning as described above was developed as a contribution to *in-vivo* multi-photon microscopy and allowed implementation of a sampling scheme named 3D custom-access serial holography (3D-CASH)[20] for random-access recordings of neuronal activity from 3D neuron circuits at high speed. In CASH measurements, the grid patterns are designed to cover the cell bodies of mammalian cortical nerve cells and are accordingly micrometer scale-sized (typically 8-15 μm). As other applications may operate on a different scale, the grid sizes and point densities are freely adaptable using the same approach. In applications using 3D samples, it is important to keep in mind that focusing of structured light will produce a 3D intensity distribution which in particular includs out-of-focus interference peaks close to the focal plane. To check the 3D intensity distribution computationally beforehand it suffices to propagate the focal light field to planes above and below the focal plane using the light propagator of Fourier optics[36]. Finally, given our findings, we estimate that AO holography offers a valuable set of features endorsing future applications in optical bioscience and possibly other areas, like material processing and metrology.

**Acknowledgement**
We wish to thank Alberto Lombardini, Cathie Ventalon, Sylvain Gigan, Jean-François Léger and Stéphane Dieudonné for helpful discussions.

**References**

1. Gabor, D. & Bragg, W. L. Microscopy by reconstructed wave-fronts. *Proceedings of the Royal Society of London. Series A. Mathematical and Physical Sciences* vol. 197 454–487 (1949).

2. Buckley, E. Holographic laser projection. *IEEE Journal of Display Technology* **7**, (2010).

3. Tahara, T., Quan, X., Otani, R., Takaki, Y. & Matoba, O. Digital holography and its multidimensional imaging applications: a review. *Microscopy* **67**, 55–67 (2018).

4. Nayak, A. R., Malkiel, E., McFarland, M. N., Twardowski, M. S. & Sullivan, J. M. A Review of Holography in the Aquatic Sciences: In situ Characterization of Particles, Plankton, and Small Scale Biophysical Interactions. *Frontiers in Marine Science* **7**, (2021).

5. Javidi, B. & Nomura, T. Securing information by use of digital holography. *Opt. Lett.* **25**, 28–30 (2000).

6. Nobukawa, T. & Nomura, T. Multilevel recording of complex amplitude data pages in a holographic data storage system using digital holography. *Opt. Express* **24**, 21001–21011 (2016).

7. Geng, Q., Wang, D., Chen, P. & Chen, S.-C. Ultrafast multi-focus 3-D nano-fabrication based on two-photon polymerization. *Nature Communications* **10**, 2179 (2019).

8. Menon, R., Patel, A., Gil, D. & Smith, H. I. Maskless lithography. *Materials Today* **8**, 26–33 (2005).

9. Desbiens, J. The Dispositif of Holography. *Arts* **8**, (2019).

10. Myung K. Kim. Principles and techniques of digital holographic microscopy. *SPIE Reviews* **1**, 1–51 (2010).

11. Papagiakoumou, E., Ronzitti, E. & Emiliani, V. Scanless two-photon excitation with temporal focusing. *Nature Methods* **17**, 571–581 (2020).

12. Jialong Chen, Zhiqiang Fu, Bingxu Chen, & Shih-Chi Chen. Fast 3D super-resolution imaging using a digital micromirror device and binary holography. *Journal of Biomedical Optics* **26**, 1–9 (2021).

13. Osten, W. *et al.* Recent advances in digital holography [Invited]. *Appl. Opt.* **53**, G44–G63 (2014).

14. Treptow, D., Bola, R., Martín-Badosa, E. & Montes-Usategui, M. Artifact-free holographic light shaping through moving acousto-optic holograms. *Scientific Reports* **11**, 21261 (2021).

15. Salomé, R. *et al.* Ultrafast random-access scanning in two-photon microscopy using acousto-optic deflectors. *Journal of Neuroscience Methods* **154**, 161–174 (2006).

16. Bullen, A., Patel, S. S. & Saggau, P. High-speed, random-access fluorescence microscopy: I. High-resolution optical recording with voltage-sensitive dyes and ion indicators. *Biophysical Journal* **73**, 477–491 (1997).

17. Reddy, G. D. & Peter Saggau. Fast three-dimensional laser scanning scheme using acousto-optic deflectors. *Journal of Biomedical Optics* **10**, 1–10 (2005).

18. Kirkby, P. A., Srinivas Nadella, K. M. N. & Silver, R. A. A compact acousto-optic lens for 2D and 3D femtosecond based 2-photon microscopy. *Opt. Express* **18**, 13720–13744 (2010).

19. Villette, V. *et al.* Ultrafast Two-Photon Imaging of a High-Gain Voltage Indicator in Awake Behaving Mice. *Cell* **179**, 1590-1608.e23 (2019).

20. Akemann, W. *et al.* Fast optical recording of neuronal activity by three-dimensional custom-access serial holography. *Nat Methods* **19**, 100–110 (2022).

21. Blochet, B., Akemann, W., Gigan, S. & Bourdieu, L. Fast wavefront shaping for two-photon brain imaging with large field of view correction. *bioRxiv* 2021.09.06.459064 (2021) doi:10.1101/2021.09.06.459064.

22. Römer, G. R. B. E. & Bechtold, P. Electro-optic and Acousto-optic Laser Beam Scanners. *Physics Procedia* **56**, 29–39 (2014).

23. Debye, P. & Sears, F. W. On the Scattering of Light by Supersonic Waves. *Proc Natl Acad Sci USA* **18**, 409 (1932).

24. Xu, J. & Stroud, R. *Acousto-Optic Devices*. (John Wiley, 1992).

25. I. C. Chang & P. Katzka. Acousto-Optic Properties of Chalcogenide Compounds. in *IEEE 1987 Ultrasonics Symposium* 511–514 (1987). doi:10.1109/ULTSYM.1987.199010.

26. VanderLugt, A. & Bardos, A. M. Design relationships for acousto-optic scanning systems. *Appl. Opt.* **31**, 4058–4068 (1992).

27. Kaplan, A., Friedman, N. & Davidson, N. Acousto-optic lens with very fast focus scanning. *Opt. Lett.* **26**, 1078–1080 (2001).

28. Reddy, G., Kelleher, K., Fink, R. & Saggau, P. Three-dimensional random access multiphoton microscopy for functional imaging of neuronal activity. *Nature Neuroscience* **11**, 713 (2008).

29. Fernández-Alfonso, T. *et al.* Monitoring synaptic and neuronal activity in 3D with synthetic and genetic indicators using a compact acousto-optic lens two-photon microscope. *Journal of Neuroscience Methods* **222**, 69–81 (2014).

30. Katona, G. *et al.* Fast two-photon in vivo imaging with three-dimensional random-access scanning in large tissue volumes. *Nature Methods* **9**, 201 (2012).

31. Akemann, W. *et al.* Fast spatial beam shaping by acousto-optic diffraction for 3D non-linear microscopy. *Opt. Express* **23**, 28191–28205 (2015).

32. Bechtold, P., Hohenstein, R. & Schmidt, M. Beam shaping and high-speed, cylinder-lens-free beam guiding using acousto-optical deflectors without additional compensation optics. *Opt. Express* **21**, 14627–14635 (2013).

33. Akondi, V. & Dubra, A. Accounting for focal shift in the Shack–Hartmann wavefront sensor. *Opt. Lett.* **44**, 4151–4154 (2019).

34. Vijay Iyer, Bradley Edward Losavio, & Peter Saggau. Compensation of spatial and temporal dispersion for acousto-optic multiphoton laser-scanning microscopy. *Journal of Biomedical Optics* **8**, 460–471 (2003).

35. Kremer, Y. *et al.* A spatio-temporally compensated acousto-optic scanner for two-photon microscopy providing large field of view. *Opt. Express* **16**, 10066–10076 (2008).

36. Goodman, J. W. *Introduction to Fourier optics*. (Roberts & Company, 2005).

37. Gerchberg, R. W. & Saxton, W. O. A Practical Algorithm for the Determination of Phase from Image and Diffraction Plane Pictures. *Optik* **35**, 237–246 (1972).

**Figure Captions**

**Figure 1 Principle of time-locked AO Bragg diffraction of femtosecond laser pulses**

**(a)** Schematic representation of Bragg diffraction of a planar incoming wavefront by acousto-optic density grating of constant frequency into an outgoing wave in first diffraction order. The input angle is tuned to the Bragg angle of the AO carrier wave. At carrier frequency this is also the output angle. For illustrative purposes angles and spatial frequency are largely exaggerated disregarding femtosecond pulse front tilt usually negligible in AODs because of the relatively small Bragg deflection angle. Positive (negative) frequency offsets will tune the output angle to positive (negative) deviations from the Bragg angle. **(b)** Positive (negative) linear frequency chirp transforms a planar input beam into a convergent (divergent) output beam. **(c)** In an AO lens the input beam is diffracted by a traveling ultrasound wave of constant positive chirp. Shown are snapshots of the lens at time zero and one cycle time later with optical ray traces in red. Direction of acoustic wave motion indicated as white arrow. While the axial position of the focus remains constant, the lateral position moves at ultrasound speed. **(d)** Two snapshots representing the lens at fractions of 0.2 and 0.8 of the AOD cycle time ($t_{cycle}$) after frequency reset (at $t_{reset}$), when the acoustic frequency is reset from its upper to its lower limit defined by the supported AO bandwidth. **(e)** Schematic representation of the synchronous mode showing the laser emission (red) together with snapshots of AO gratings on a time scale relative to the AOD cycle time for a series of consecutive switches between gratings at constant carrier frequency (green), with negative frequency chirp (blue), positive frequency offset (red), with positive frequency chirp (yellow) and negative frequency offset (cyan). **(f)** Illustration of focus positions in the y-z plane in the image space of a Fourier lens receiving wavefronts shaped by the AO gratings in (e). **(g)** Optical conjugation of two crossed AO lenses of identical focal length (same AO frequency chirp) provides spherical phase modulation on both axes which adds up to give a spherical lens. **(h)** Illustration of AO gratings (*left*) and the resulting phase modulations of the output beam (*middle, right*) by two crossed two AODs (X- and Y-AOD) for the purpose of 3D scanning of the input beam. **(i)** Orthogonal polynomial phase functions with biaxial symmetry compatible with the crossed AOD configuration, up to $8^{th}$ polynomial order in scaled color image representation. For every order there are two functions. The polynomial order is indicated on top of each pair.

**Figure 2 Holographic light patterning by acousto-optic frequency modulation**

**(a)** Illustration of wavefront shaping by frequency-modulated acousto-optic diffraction gratings. **(b)** Illustration of the input beam transformation performed by a Fourier lens, for instance a microscope objective, which optically transforms the light field in the back focal plane of the lens into the resulting light field in the front focal plane. Mathematically, the transformation is performed by fast Fourier transformation (FFT) of the input field with A, the amplitude, and φ, the phase. **(c)** Schematic representation of the Gerchberg-Saxton iterative Fourier transform algorithm with adaptive target function. $A_{x/y}$ and $\varphi_{x/y}$ are the 1D amplitude and phase distributions aligning either with the x- or the y-axis as the active axes in the dual X-Y-AOD configuration. IFFT designates the inverse Fourier transformation. The resulting phase hologram is labeled as 'Hologram'. **(d)** Results obtained with the 1D iterative Fourier algorithm (c) for a series of targets of 1, 3 5 and 7 equidistant points (gray) showing the hologram (blue), the associated FM signal (red) and the 1D reconstruction of the hologram (black). L denotes the overall length of the target. **(e)** *Top row:* 2D phase holograms derived by assigning the same 1D solutions after unwrap to both axes (x,y). *Middle row:* 3x3, 5x5 and 7x7 patterns of focal amplitude reconstructed from the holograms given above. *Bottom row:* same for the intensity. **(f)** Example different phase holograms for x- and y-axis giving a 5x3 focal spot pattern. **(g)** Same as (f), but higher spatial frequency of the y-axis hologram ($\varphi_y$). **(h)** Example of a 5x4 point pattern with unequal distance between points in x- and y-direction.

**Figure 3 Acousto-optic holography with amplitude modulation**

**(a)** 1D Fourier Transform algorithm with a static target function. As before (Fig. 2c), 'Hologram' designates phase hologram. **(b)** Results obtained from (a) for the same multi-point target functions (*left column*; gray) as before (Fig. 2d), but used as static targets, with reconstructions (*middle column*; black) of the phase hologram alone (FM) and combined phase and amplitude hologram (FM/AM). The holograms are shown as overlay (*right column*) of amplitude (black) and phase (red). **(c)** 2D amplitude and phase holograms derived by applying the 1D holograms in (b) to both axes (x,y). **(d)** Focal amplitude (*left*) and phase (*middle*) reconstructions of the holograms in (c) using the phase alone (FM) or together with the amplitude (FM/AM), together with the spatial variance of the reconstructed focal phase (*right*) in FM and FM/AM mode. **(e)** Reconstructions of the 1D hologram of the five-point target (b) at variable AM gain, showing the apodization signal (*top*; blue) and the reconstructed amplitude (*bottom*; black). **(f)** Target error (*top*), evaluated as deviation of the amplitude of the reconstructed and imposed target peaks and the target peaks, as function of AM modulation gain, together with the AOD power transmission (*bottom*) calculated for an input beam of flat intensity (blue).

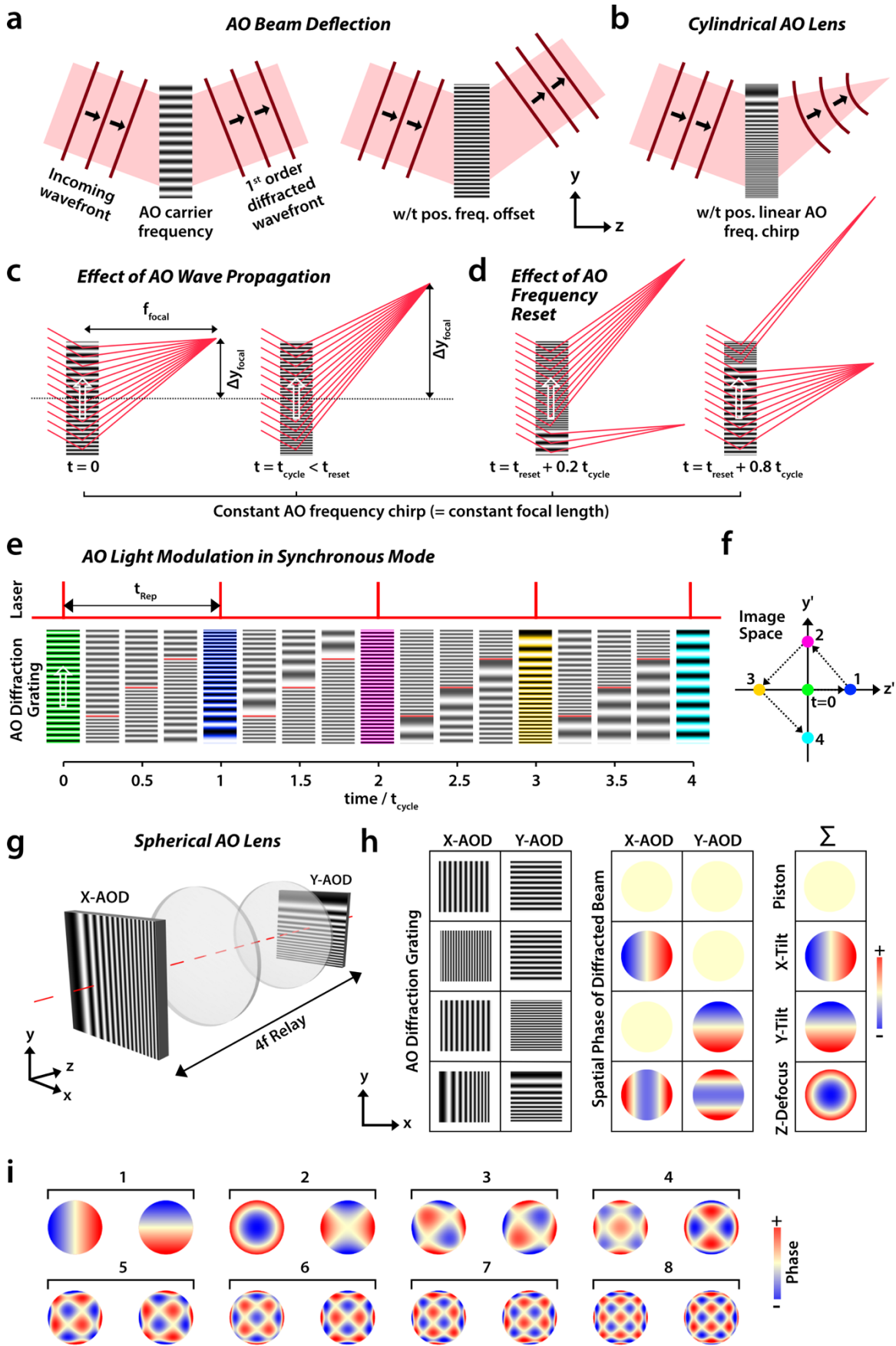

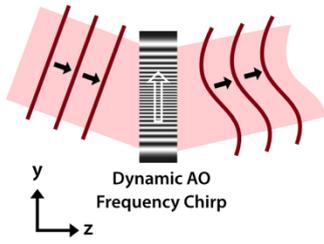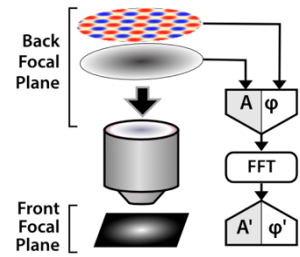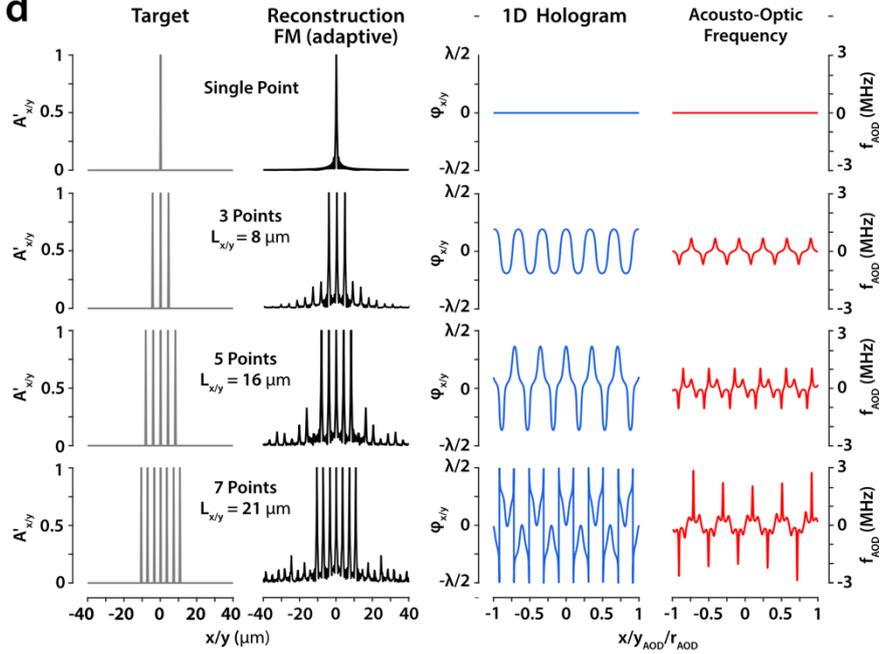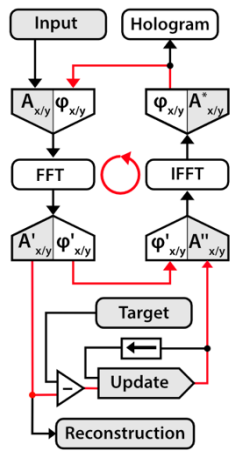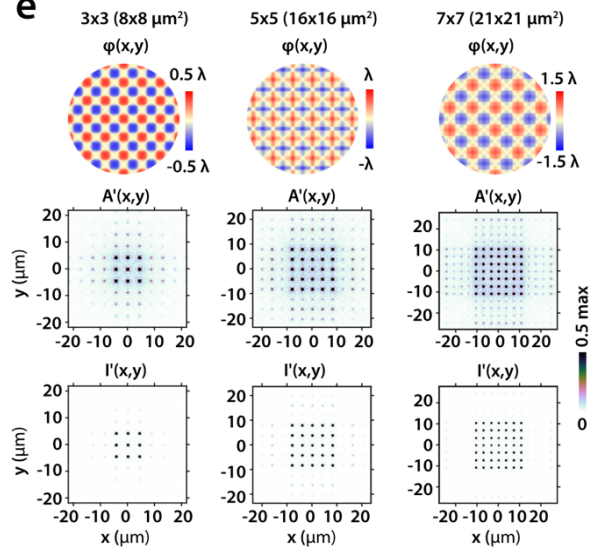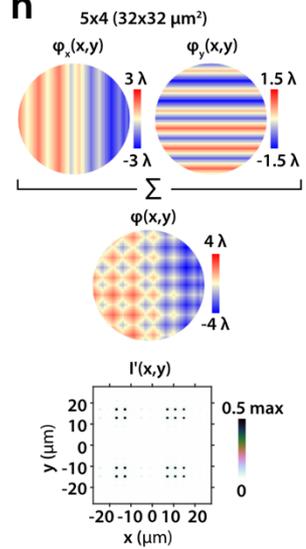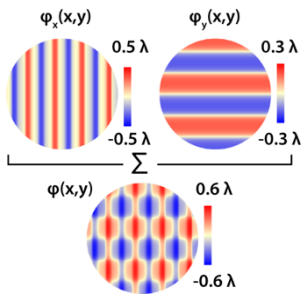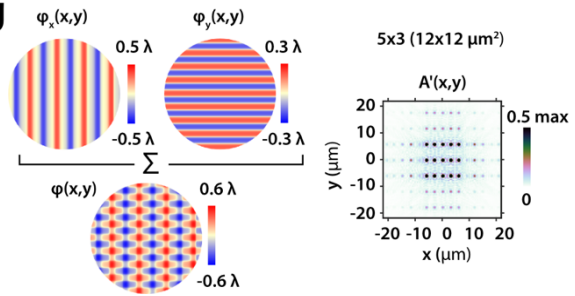

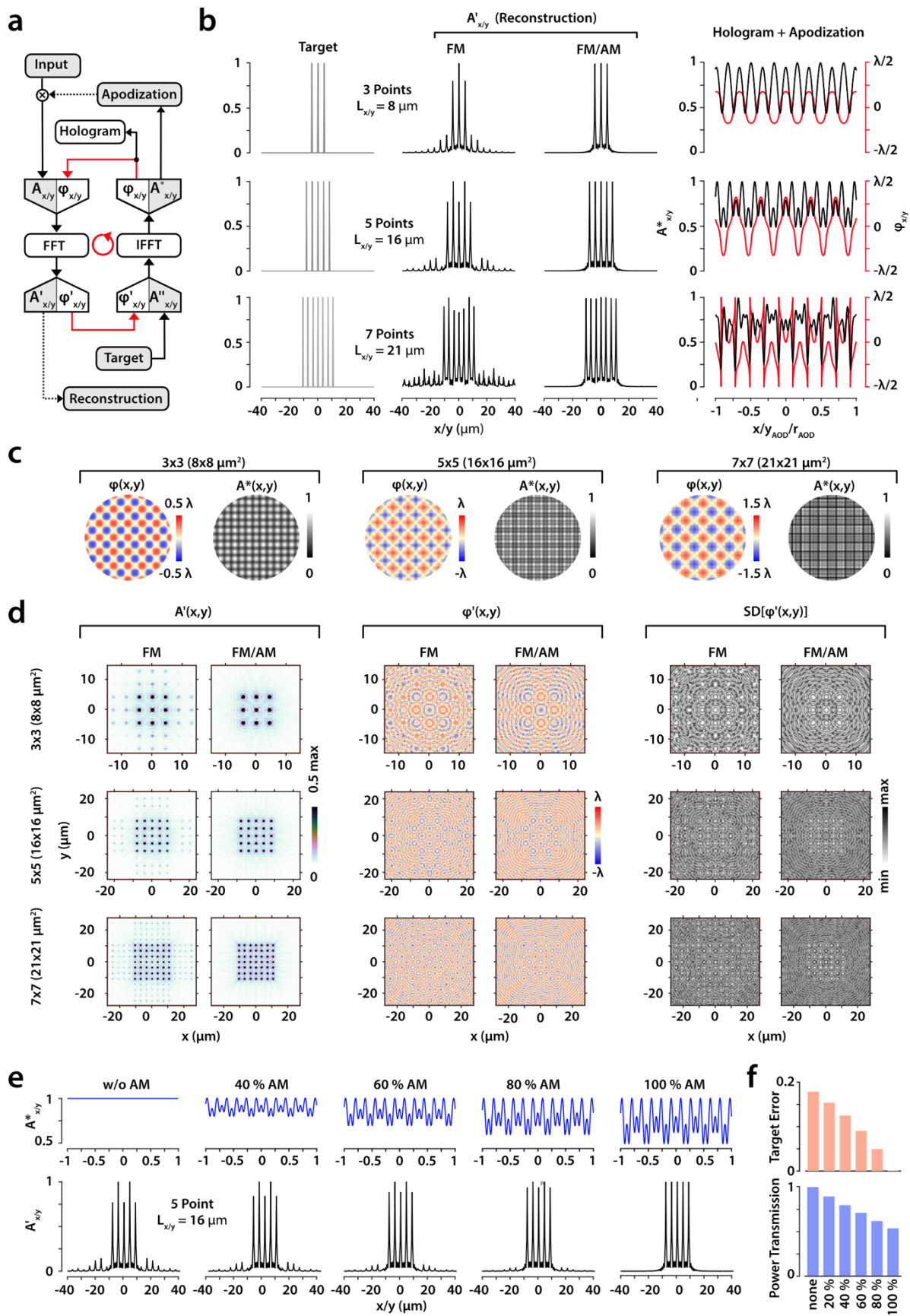